\documentclass[twocol]{ametsoc}

\journal{jamc}

\bibpunct{(}{)}{;}{a}{}{,}

\title{Tropical cyclone size is strongly limited by the Rhines scale: experiments with a barotropic model\\
{\color{red}NOT PUBLISHED. Submitted for peer review}}

    \authors{Kuan-Yu Lu and Daniel R. Chavas}

    \affiliation{Purdue University, Department of Earth, Atmospheric, and Planetary Sciences, West Lafayette, IN} 

\abstract{Recent work found evidence using aquaplanet experiments that tropical cyclone size on Earth is limited by the Rhines scale, which depends on the planetary vorticity gradient, $\beta$.
This study aims to examine how the Rhines scale limits the size of an individual tropical cyclone.  The traditional Rhines scale is first re-expressed as a vortex Rhines scale and Rhines speed to characterize how wave effects vary with radius in a vortex whose wind profile is known. Experiments are performed using a simple barotropic model on a $\beta$-plane initialized with a TC-like axisymmetric vortex defined using a recently-developed theoretical model for the tropical cyclone wind profile. $\beta$ and initial vortex size are each systematically varied to investigate the detailed responses of the TC-like vortex to $\beta$. Results show that the vortex shrinks towards an equilibrium size that closely follows the vortex Rhines scale. Physically, this scale divides the vortex into a vortex-dominant region at small radii, where the axisymmetric circulation is steady, and a wave-dominant region at larger radii, where the circulation stimulates Rossby waves and dissipates. A larger initial vortex relative to its vortex Rhines scale will shrink faster, and the shrinking timescale is well described by the vortex Rhines timescale, which is defined as the overturning timescale of the circulation at the vortex Rhines scale and is shown to be directly related to the Rossby wave group velocity. 
The relationship between our idealized results and the real Earth is discussed.} 

\begin{document}

\maketitle

%
%
\textbf{Significance Statement}
Tropical cyclones vary in size significantly on the Earth, but how large a tropical cyclone could potentially be is still not understood. This study derives a new parameter, called the "vortex Rhines scale", and shows in a simple model how it can predict the upper limit on tropical cyclone size and describe how fast a tropical cyclone larger than this size will shrink. These results help explain why tropical cyclone size tends to increase slowly with latitude on Earth.
%

%
\section{Introduction}

The size of a tropical cyclone (TC) determines its footprint of gale-force winds \citep{Powell_and_Reinhold_2007}, storm surge \citep{Irish_et_al_2011} and rainfall \citep{Kidder_et_al_2005,Lavender_and_McBride_2020}. Therefore, understanding the dynamics of TC size is important for understanding potential TC impacts. 

Observational studies have found that TC size can vary significantly. For example, the TC radius of vanishing wind ($R_0$) typically ranges from $400$km to $1100$km \citep{Chavas_et_al_2016}. Past studies have shown that TC size may be sensitive to a variety of parameters, such as synoptic interaction \citep{Merrill_1984,Chan_and_Chan_2013}, time of the day \citep{Dunion_et_al_2014}, environmental humidity \citep{Hill_and_Lackmann_2009}, and latitude \citep{Weatherford_and_Gray_1988a,Weatherford_and_Gray_1988b,Chavas_et_al_2016}.
On the f-plane, TC size decreases with decreasing Coriolis parameter $f$ following an $f^{-1}$ scaling \citep{Chavas_and_Emanuel_2014,Khairoutdinov_and_Emanuel_2013,Zhou_et_al_2014}, suggesting that size should decrease rapidly with latitude in the tropics. However, in observations, TCs size tends to increase slowly with latitude \citep{Kossin_et_al_2007,Knaff_et_al_2014}. Indeed, \cite{Chavas_et_al_2016} showed explicitly that the inverse-f dependence can not explain the observed dependence of TC size with latitude. 

Recently, \citet[][hereafter CR19]{Chavas_and_Reed_2019} used aquaplanet experiments with uniform thermal forcing to demonstrate that median TC size scales with the Rhines scale \citep{Rhines_1975}. This scale depends inversely on the planetary vorticity gradient, $\beta$, and increases very slowly with latitude in the tropics, which matches the behavior seen in observations. Their findings lead to the question: is the size of an \textit{individual} storm limited by the Rhines scale and, if so, why? 

The Rhines scale has traditionally served as the scale that divides flow into turbulence and Rossby wave-dominated features \citep{Held_and_Larichev_1996}. When the eddy length scale is at or larger than the Rhines scale, and the linear Rossby wave term dominates the nonlinear turbulent term. Conceptually, this implies that an eddy circulation larger than the Rhines scale behaves more wave-like. Despite being used to understand the role and scale of eddies in many large-scale atmospheric and oceanic circulation theories \citep[e.g.,][]{James_and_Gray_1986, Vallis_and_Maltrud_1993,Held_and_Larichev_1996,Held1999,Lapeyre_and_Held_2003,Schneider_2004,LaCasce_and_Pedlosky_2004}, it has not been applied to understand the scale of the tropical cyclone. Since a TC can be regarded as an eddy circulation embedded in the tropical atmosphere, it seems plausible that the Rhines scale can indeed directly modulate TC size. However, it is not clear how $\beta$ acts to limit the size of an individual TC and how this may be understood in the context of the Rhines scale.

The Rhines scale is typically calculated using a single characteristic turbulent velocity, usually defined as the root-mean-square velocity \citep[e.g.,][]{Sukoriansky_et_al_2006,Kidston_et_al_2010}, or even a single characteristic velocity scale for a TC (an outer circulation velocity scale $U_{\beta}$ in CR19; a collective velocity scale in \citealt{Hsieh_et_al_2020}). However, a TC clearly does not possess a single velocity scale but instead has rotational velocities that vary strongly as a function of radius. Moreover, a theoretical model now exists for the radial structure of the TC wind field that captures the first-order behavior of TC structure found in observations \citep{Chavas_et_al_2015,Chavas_and_Lin_2016}. Such a wind field model may be used to understand the detailed dynamics of this Rhines scale effect within a TC.

The simplest way to test the effect of $\beta$ on a TC-like vortex is to perform barotropic model simulations of a single TC-like vortex on a $\beta$-plane, as the low-level circulation of a TC may be considered approximately barotropic and the Rhines scale arises from the barotropic vorticity equation itself. This approach neglects the role of the secondary circulation, which is undoubtedly an important part of TC dynamics, in order to isolate the basic behavior and dynamical response of the primary circulation of a TC. Various aspects of the dynamics of a vortex on a $\beta$-plane have been analyzed in past studies. The most widely-known effect is $\beta$-drift \citep{Chan_and_Williams_1987}, which is the poleward and westward vortex translation induced by the interaction of the vortex and the vortex-generated planetary Rossby waves\footnote{We use the term ``planetary'' to emphasize that these Rossby waves arise due to vortex flow across the meridional vorticity gradient of the Earth's rotation, $\beta$. This is in contrast to ``vortex Rossby waves'' [\citealt{Montgomery_and_Kallenbach_1997}], which are Rossby waves that arise from vortex flow across the radial relative vorticity gradient of the vortex itself.} \citep{Llewellyn_and_Smith_1997,Sutyrin_and_Flierl_1994,Fiorino_and_Elsberry_1989,Wang_et_al_1997}. Notably, though not emphasized in their study, \citet{Chan_and_Williams_1987} demonstrated in their $\beta$-drift experiments that vortex size tends to decrease with time on a $\beta$-plane. While inducing translation, these Rossby waves transfer kinetic energy from vortex to Rossby waves that then propagate into the environment \citep{Flierl_and_Haines_1994,Sutyrin_et_al_1994,Smith_et_al_1995}, thereby weakening the primary circulation and hence reducing vortex size \citep{McDonald_1998,Lam_and_Dritschel_2001,Eames_and_Flor_2002}. \citet{Eames_and_Flor_2002} found that for larger vortex size, the Rossby wave generation will dominate the dynamics of the vortex and, further, the vortex translation speed is correlated with the Rossby waves phase speed, a result that is conceptually similar to the Rhines scale effect described above. However, past work has yet to systematically test and explain the response of the size of an individual TC to $\beta$ and place it in the context of the Rhines scale.

Here we focus on understanding the detailed response of the structure and size of an individual TC-like vortex to $\beta$.  Our principal research questions are:
\begin{enumerate}
    \item How does the size of and individual TC-like vortex respond to $\beta$?
    \item Can we develop a framework explaining why $\beta$ limits storm size and its relationship to the traditional Rhines scale?
    \item Can we predict the time-dependent response of TC size to $\beta$?
\end{enumerate}
To answer these questions, we first revisit the meaning of the Rhines scale in the context of an individual coherent vortex and show how it may be re-expressed in the context of a vortex with a known wind profile that is useful for understanding the response of an axisymmetric vortex to $\beta$. We then use this theory to analyze dynamical experiments using a simple barotropic model on a $\beta$-plane initialized with the axisymmetric low-level tropical cyclone wind field defined in \citet{Chavas_et_al_2015}. We conduct experiments systematically varying $\beta$ and initial vortex size to investigate the detailed time-dependent response of the vortex to the $\beta$. Overall, our focus is on understanding the nature of the response of the size of a TC-like vortex to $\beta$ and how we may use the conceptual foundation of the Rhines scale to predict it. Analysis of the details of energy transfer between the vortex and Rossby waves is left for future work.

The paper is organized as follows: Section 2 presents the theory and proposes our hypotheses; Section 3 demonstrates our model configuration and experiment designs; Section 4 presents the idealized results and analyses of our experiments; Section 5 presents our key findings and discusses the implications of our results and their relationship to real TCs on Earth.

\section{Theory}

The goal of this study is to investigate the limitation of vortex size by the Rhines scale, which is a parameter that governs the interaction between Rossby waves and a vortex in a fluid. This effect exists in any fluid in the presence of a planetary vorticity gradient, $\beta = \frac{\partial f}{\partial y}$, where $f$ is the Coriolis parameter and $y$ is the meridional direction. The simplest such system is a dry barotropic (i.e. single-layer) fluid with constant depth. Such a fluid will obey the non-divergent dry barotropic vorticity equation as following:
\begin{equation}\label{eq:BVE}
\underbrace{\frac{\partial \zeta}{\partial t}}_\text{Tendency term} = \underbrace{-\overrightarrow{\textbf{u}} \cdot \bigtriangledown \zeta}_\text{Non-linear advection term} \underbrace{-\beta v}_\text{Beta term},
\end{equation}
where $\partial$/$\partial t$ is the Eulerian tendency, $\zeta$ is the relative vorticity, $\overrightarrow{\textbf{u}}$ is horizontal wind velocity, $v$ is the meridional wind speed, and $\beta$ is the meridional gradient of planetary vorticity. The term on the left hand side is the vorticity tendency, the first RHS term is the non-linear advection of relative vorticity (hereafter "non-linear term"), and the second RHS term is the linear advection of the planetary vorticity (hereafter "$\beta$ term"). Note that, unlike a shallow-water system, this system has no gravity waves because the fluid depth is constant.

To determine which term in the barotropic vorticity equation dominates the vorticity tendency, a traditional scale analysis of Eq. \ref{eq:BVE} would yield:
\begin{equation}\label{eq:Ori_scaleBVE}
\frac{V}{LT} = -\frac{V^2}{L^2} - \beta V,
\end{equation}
where the $V$ is the speed scale of the wind, $L$ is the horizontal length scale, $T$ is the time scale. For a non-divergent axisymmetric vortex, $\overrightarrow{\textbf{u}}$ and $v$ can be both expressed as the azimuthal wind speed of the vortex circulation $U_c$. 
Note that the advection term in Eq. \ref{eq:Ori_scaleBVE} has $L^2$ in denominator, but each $L$ has a different physical meaning: the relative vorticity, $\zeta = \partial(rU_c)/dr$, represents the radial gradient of $U_c$, and hence $L_{\zeta} = R$; in contrast, the advection operator, $\overrightarrow{\textbf{u}} \cdot \bigtriangledown$, is the tangential advection around the circumference of the vortex, and hence $L_{\overrightarrow{\textbf{u}} \cdot \bigtriangledown} = 2 \pi R$. Together, in Eq. \ref{eq:Ori_scaleBVE}, the denominator becomes $L^2 = 2 \pi R^2$. Therefore, the ratio between the non-linear advection term and the $\beta$ term, which we define as the Rhines number ($Rh$), can be written as following:

\begin{equation}\label{eq:TermRatio}
\frac{\overrightarrow{\textbf{u}} \cdot \bigtriangledown \zeta}{\beta v} \equiv Rh \approx \frac{\frac{U_c^2}{2 \pi R^2}}{\beta U_c}=\frac{U_c}{2 \pi \beta R^2}.
\end{equation}
Eq. \ref{eq:TermRatio} neglects any radial flow as is required for a barotropic vortex. Real TCs possess significant inflow at low-levels, which is a topic we address in the discussion of our results below.

When $Rh=1$, the non-linear term and the $\beta$-term are equal \citep[][p.446]{Vallis_2017_p.446}. We can rearrange Eq. \ref{eq:TermRatio} to define the Rhines scale ($R_{Rh}$) for a rotational flow with speed of $U_c$:
\begin{equation}\label{eq:RhinesScale}
R_{Rh} \equiv \sqrt{\frac{{U}_{c}}{2 \pi \beta}}.
\end{equation}

Note that the key distinction of Rhines scale from a deformation-type scale is that the velocity scale is a true flow speed rather than a gravity wave phase speed. We emphasize that we are precise in including the $2\pi$ factor in our explanation above, as it is quantitatively important for our results presented below. This is in contrast to typical scale analyses which are agnostic to the inclusion or neglect of constant factors (and indeed the appearance of such factors varies in the literature). 

\begin{figure*}[t]
 \centerline{\includegraphics[width=38pc]{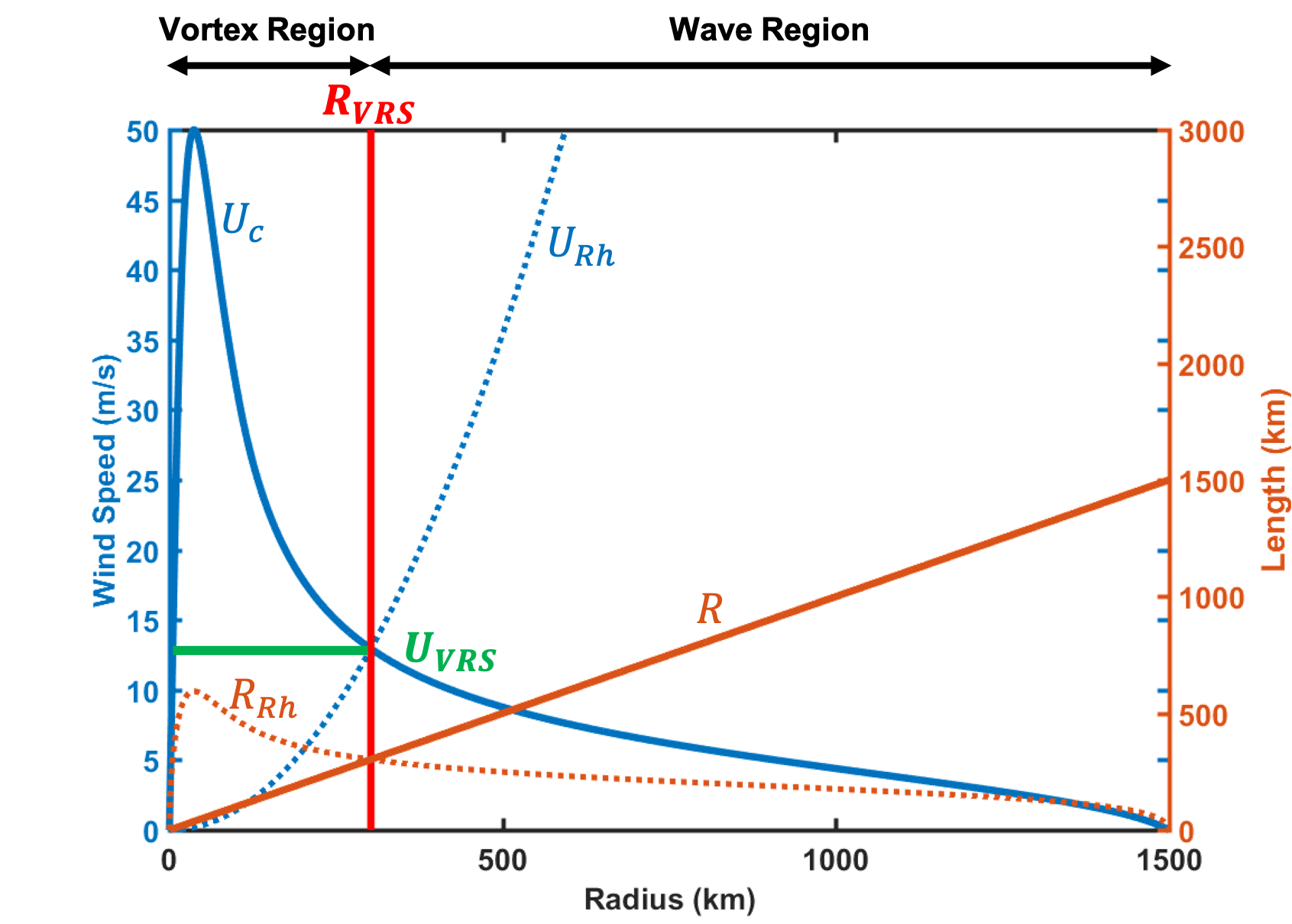}}
 \caption{Conceptual diagram of the vortex Rhines scale ($R_{VRS}$, red vertical line) defined by the vortex's tangential wind profile ($U_c$, blue solid line) and the Rhines speed profile ($U_{Rh}$, eq.\ref{eq:RhinesSpeed}, blue dashed line). Brown dashed line represents the Rhines scale at each radius ($R_{Rh}$, eq.\ref{eq:RhinesScale}), and green horizontal line indicates the vortex Rhines speed ($U_{VRS}$). Regions with different dynamical features are labeled, inside $R_{VRS}$ is the vortex region, and outside $R_{VRS}$ is the wave region.}
 \label{fig:Conceptual_Diagram}
\end{figure*}
 
Previous studies tend to estimate a system's Rhines scale by assuming a characteristic eddy flow velocity \citep[e.g., CR19;][]{Chemke_and_Kaspi_2015,Frierson_et_al_2006}, such that the system will have a single characteristic Rhines scale. However, an individual TC-like vortex possesses circulation speeds that vary strongly with radius and hence contains both "small" and "large" circulations simultaneously by definition. Thus, the details of these wave effects should depend on radius and cannot be characterized by a single velocity nor a single Rhines length scale. The simplest starting point is to consider the vortex as comprised of independent circulations at each radius with different wind speeds and calculate their corresponding Rhines scale as a function of radius. This results in a radial profile of $R_{Rh}$ that depends on the vortex's tangential wind profile and $\beta$ (eq. \ref{eq:RhinesScale}). Here we use the wind speed of the TC's circulation ($U_c$), which varies with radius, to calculate the radial profile of the Rhines scale in a TC. This allows one to evaluate how $\beta$ may affects the vortex circulation differently at different radii. 

For a sufficiently large circulation relative to the Rhines scale ($R \gg R_{Rh}$), the non-linear term is small ($Rh \ll 1$) and thus its vorticity tendency will be governed by the $\beta$ term, which gives pure Rossby waves. Therefore, a circulation at a given radius that is larger than its corresponding Rhines scale will be affected by $\beta$ and generate Rossby waves, which will radiate energy away.
Meanwhile, for a sufficiently small circulation relative to the Rhines scale ($R \ll R_{Rh}$), the $\beta$ term is small ($Rh \gg 1$), and thus its vorticity tendency will be governed by the non-linear term and will generate minimal Rossby wave activity. For a perfectly axisymmetric vortex, the relative vorticity advection term is zero, and hence the circulation simply circulates without the energy sink from Rossby waves.

The concept of the Rhines scale for a vortex can be more easily understood if rephrased in terms of a velocity scale rather than length scale. 
Since $R_{Rh}$ is function of $U_c$ (eq.\ref{eq:RhinesScale}), that means for a given wind speed, $R_{Rh}$ defines a specific length scale whose value relative to the circulation radius determines whether the circulation will generate significant Rossby waves or not. Alternatively, one may choose to fix the circulation radius, in which case $Rh$ equivalently defines a specific wind speed whose value relative to the wind speed at radius determines whether the circulation will be affected by Rossby waves or not.
Based on this concept, we can rearrange the Eq. \ref{eq:TermRatio} to yield the wind speed when $Rh=1$, which we define as the Rhines speed ($U_{Rh}$): 
\begin{equation}\label{eq:RhinesSpeed}
{U}_{Rh}\equiv 2 \pi \beta {R}^2
\end{equation}
Importantly, the radial dependence of $U_{Rh}$ is solely a function of $\beta$, and for a given value of $\beta$, its profile is fixed and is independent of the vortex. The case $U_{c} \ll U_{Rh}$ is analogous to $R \gg R_{Rh}$ and corresponds to the wave-dominant regime, while the case $U_{c} \gg U_{Rh}$ is analogous to $R \ll R_{Rh}$ and corresponds to the vortex-dominant regime.

The above definition is especially convenient because the radial structure of the circulation may be known or specified. Figure \ref{fig:Conceptual_Diagram} displays an example radial profile of $U_{c}$ (blue solid line) for a tropical cyclone, defined by the model for the low-level azimuthal wind of Chavas et al. (2015; described in detail in Section 3 below), and the $U_{Rh}$ (blue dashed line) and $R_{Rh}$ (brown dashed line) that are calculated using Eqs. \ref{eq:RhinesSpeed} and \ref{eq:RhinesScale}, respectively.
Typically, a TC-like $U_{c}$ profile will decrease monotonically with radius outside the radius of maximum wind. Meanwhile, the $U_{Rh}$ (blue dashed line) profile increases with radius monotonically. Therefore, there will be an intersection between these two curves at a specific radius, which we define as the vortex Rhines scale (${R}_{VRS}$, red vertical line). For the convenience of later discussion and analyses, we also define the value of $U_c$ at $R_{VRS}$ as the vortex Rhines speed ($U_{VRS}$, green horizontal line). 
We may further define the turn-over time scale of the circulation at $R_{VRS}$ as the vortex Rhines timescale ($T_{VRS}$), which can be written as:
\begin{equation}\label{eq:TVRS}
T_{VRS} = \frac{2{\pi}R_{VRS}}{U_{VRS}},
\end{equation}

Analogous to the traditional Rhines scale separating wave and vortex dominant regimes, ${R}_{VRS}$ divides the vortex into a vortex-dominant region at smaller radii and a wave-dominant region at larger radii, as shown by Figure \ref{fig:Conceptual_Diagram}, which we define as the "wave region" and "vortex region".
In the vortex region at smaller radii, planetary Rossby waves are not readily generated and the rapid rotation will axisymmetrize the vorticity field \citep{Montgomery_and_Kallenbach_1997} until the vortex flow is parallel to the vorticity contours and the flow become quasi-steady (Eq. \ref{eq:BVE}).
Meanwhile, in the wave region at larger radii, the rotating flow is slow enough to generate significant planetary Rossby wave activity, which will cause asymmetric deformation of the vortex flow at those radii. 
Taken together, the expectation is that only the circulation in wave-dominant region will stimulate significant Rossby waves, distorting and gradually dissipating the circulation therein. Meanwhile, the circulation in the vortex-dominant region would be expected to remain nearly steady. As a result, vortex size will be limited by ${R}_{VRS}$.

To investigate how ${R}_{VRS}$ affects the size of a TC-like vortex, we propose following hypotheses: 
\begin{enumerate}
  \item Vortex size is limited by its vortex Rhines scale, ${R}_{VRS}$.
  \item A larger/smaller vortex relative to its vortex Rhines scale will shrink faster/slower.
\end{enumerate}
Below we test these hypotheses by simulating a TC-like vortex on a $\beta$-plane. We focus here on characterizing and understanding the vortex response to $\beta$ across experiments varying TC size and $\beta$. Theoretical analysis of the detailed energetics of this response is left to future work.

\begin{figure}[t]
 \centerline{\includegraphics[width=19pc]{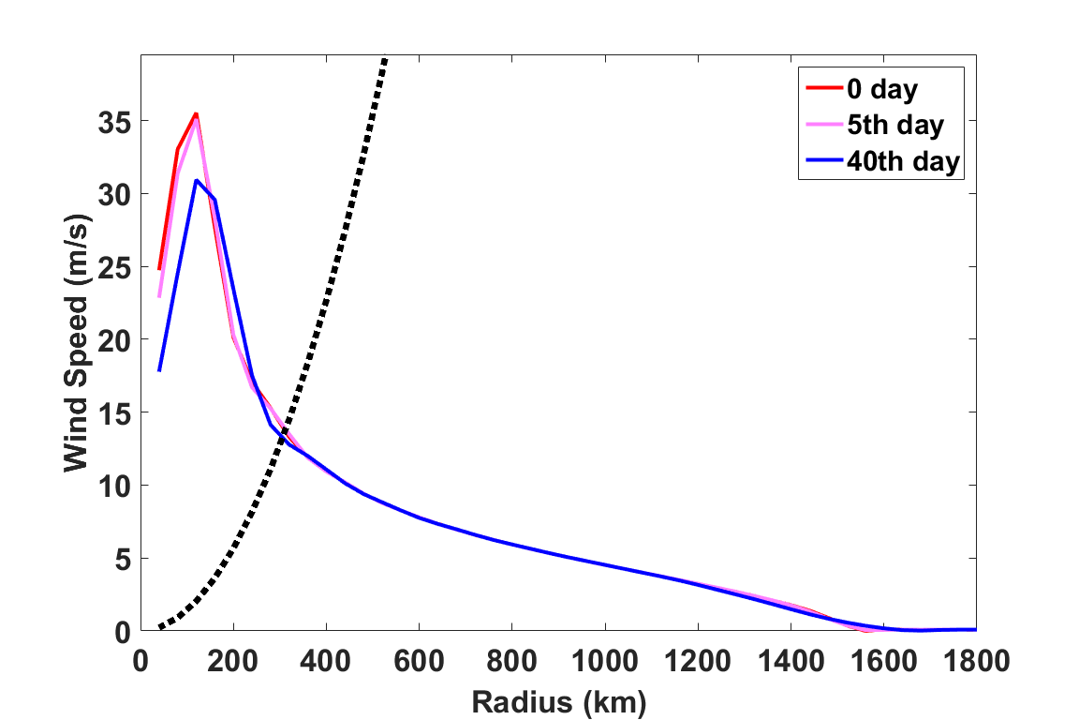}}
 \caption{Evolution of the $CTRL$ azimuthal-mean tangential wind profile on the $f$-plane ($\beta = 0$) at day 0, 5, and 40.}
  \label{fig:FPlane}
\end{figure}

\section{Methods}

\subsection{Barotropic model} 
This study uses a non-divergent, dry barotropic model to simulate the vortex behavior on a $\beta$-plane.
We use the open-source model developed by James Penn and Geoffrey K. Vallis (available at: http://empslocal.ex.ac.uk/people/staff/gv219/codes/barovort.html). It uses a pseudo-spectral method with double-periodic boundary conditions to solve the barotropic vorticity equation (Eq. \ref{eq:BVE}) in 2-D space.
The model is set up with 500 grid points in both $x$ and $y$ directions, with grid spacing of 20-40 km depending on the experiment. The initial time-step is 60 seconds and an adaptive time-step is used thereafter to avoid violating the CFL condition. The forcing amplitude factor and the dissipation term are both set as zero. For numerical stability, the model applies a high wave-number cutoff to damp wave-numbers larger than 30. Experiments with different wave-number threshold are tested (40, 50, and 60) and results indicate that they are not sensitive to different threshold (not shown).

There are two principal advantages of using such a simple barotropic model. First, since the barotropic vorticity equation only includes relative and planetary vorticity advection terms, it is an ideal tool to isolate the dynamical details of the effects of the Rhines scale on a TC-like vortex, which is directly defined by the ratio of these two terms. Second, the barotropic model is non-divergent and hence neglects the boundary layer inflow and upper level outflow found in a real TC. As a result, radial momentum transport across radii is neglected, and hence the simulated vortex's circulation at each radius is nearly independent, which simplifies understanding of the dynamical response.

\subsection{Tropical cyclone wind field model} 

Since our interest is in tropical cyclones, we employ the model of \citet[][, hereafter C15 model]{Chavas_et_al_2015} for the complete radial profile of the TC low-level azimuthal wind field to initialize the barotropic model. C15 model is a theoretical model that can reproduce the first-order structure of the TC wind field and also dominant modes of wind field variability \citep{Chavas_and_Lin_2016}. The model wind profile may be specified by a small number of storm and environmental physical parameters. The storm parameters are the maximum wind speed ($V_{max}$), the outer radius of vanishing wind ($R_0$), and the Coriolis parameter, $f$; the environmental parameters are the radiative-subsidence rate ($w_{cool}$) and surface drag coefficient ($C_d$) for the outer region and the ratio of surface coefficients of enthalpy and drag ($C_k/C_d$). In this study, we fixed $w_{cool}=0.002$ $m/s$, $C_d=0.0015$, $C_k/C_d = 1$, $C_{dvary}=0$, $C_kC_{dvary}=0$, ${eye}_{adj}=0$, and $\alpha_{eye}=0.15$.
We set the Coriolis parameter to be constant at its value at $10^{\circ}$N in order to keep our initial wind profile fixed with respect to $f$, including for our experiments varying $\beta$ below. On the sphere this would not be possible since $f$ and $\beta$ both depend on latitude. Here though we seek to isolate the effect of varying $\beta$ alone, which we can conveniently be done in a $\beta$-plane model since $f$ does not appear in the governing equation at all.
Note that for the set of input parameters given above, the C15 model will implicitly predict the radius of maximum wind ($R_{max}$). To input the vortex into the barotropic model, the wind profile is transformed into an axisymmetric vorticity field and placed at the domain center to define the barotropic model's initial condition.

\begin{figure*}[t]
 \centerline{\includegraphics[width=38pc]{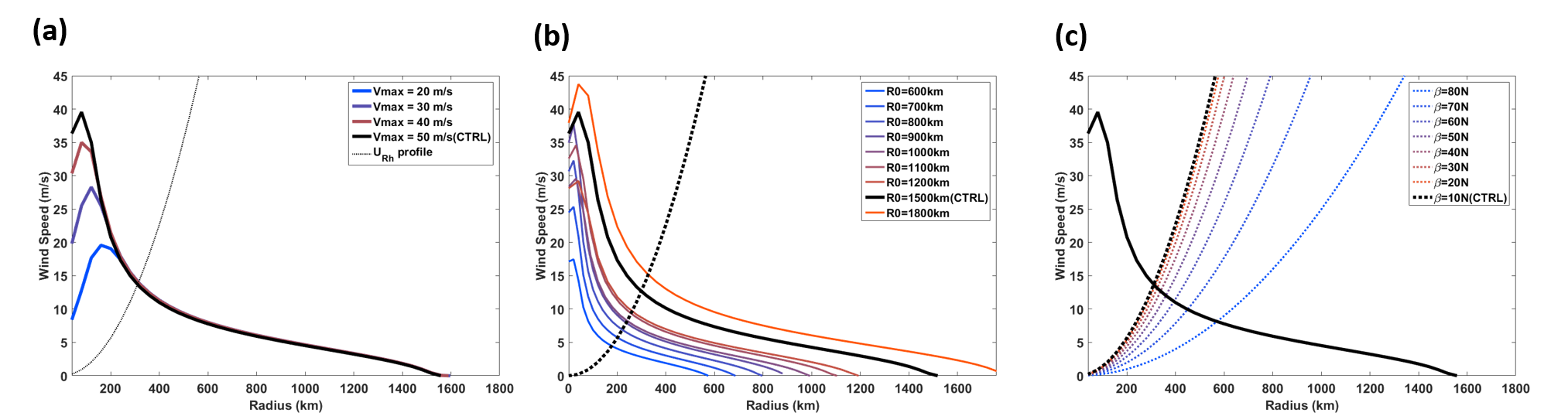}}
 \caption{Initial azimuthal-mean tangential wind profiles ($U_c$) and Rhines speed profiles ($U_{Rh}$) for all experiments within each of our three experiment sets. (a) $VARYVMAX$, varying initial intensity; black dashed line represent $U_{Rh}$ profile. (b) $VARYR0$, varying initial outer size. (c) $VARYBETA$, varying $\beta$, with $U_{Rh}$ profiles in colored dashed lines. The $CTRL$ profile is represented by the black solid curve across all plots.}
  \label{fig:Initial_Condition}
\end{figure*}

\begin{table}[h]
\caption{Parameter values for experiments in each experiment set described in the text.}\label{t1}
\begin{center}
\begin{tabular}{ccccc}
\topline
                     & R0(km) & $\beta$ & Vmax (m/s) & Grid Space (km) \\
\midline
VARYR0               & 600    & 10N  & 50         & 20            \\
                     & 700    & 10N  & 50         & 20            \\
                     & 800    & 10N  & 50         & 20            \\
                     & 900    & 10N  & 50         & 20            \\
                     & 1000   & 10N  & 50         & 30            \\
                     & 1100   & 10N  & 50         & 30            \\
                     & 1200   & 10N  & 50         & 40            \\
                     & 1500   & 10N  & 50         & 40            \\
                     & 1800   & 10N  & 50         & 40            \\
\midline
VARYBETA             & 1500   & 10N  & 50         & 40            \\
                     & 1500   & 20N  & 50         & 40            \\
                     & 1500   & 30N  & 50         & 40            \\
                     & 1500   & 40N  & 50         & 40            \\
                     & 1500   & 50N  & 50         & 40            \\
                     & 1500   & 60N  & 50         & 40            \\
                     & 1500   & 70N  & 50         & 40            \\
                     & 1500   & 80N  & 50         & 40            \\
\midline
VARYVMAX             & 1500   & 10N  & 20         & 40            \\
                     & 1500   & 10N  & 30         & 40            \\
                     & 1500   & 10N  & 40         & 40            \\
                     & 1500   & 10N  & 50         & 40            \\
\botline
\end{tabular}
\end{center}
\end{table}

\subsection{Experiments} 

The initial wind profiles for all experiment sets described below are displayed in Figure \ref{fig:Initial_Condition}, with parameters listed in Table 1.
\subsubsection{Control Experiment}

We define our Control experiment ("$CTRL$") as a simulation with a uniform quiescent environment and a single vortex at the domain center. $\beta$ is fixed at a value of $2.2547 \times 10^{-11}$ $m^{-1}s^{-1}$ corresponding to a latitude of $10^{\circ}$ N. The vortex has $V_{max}=50 ms^{-1}$ and $R_0=1500km$. The $CTRL$ is used below to illustrate the basic vortex response on a low latitude $\beta$-plane.

In our experiment framework, when inserting a TC-like vortex into the $f$-plane (setting $\beta = 0$), the tangential wind profile will exhibit an initial adjustment at small radii but then will remain very steady for many tens of days as shown in Fig. \ref{fig:FPlane}. Despite the initial inner-core structural adjustment, the outer circulation remains nearly unchanged from its initial state. This result demonstrates that the vortex circulation is spun up and the vortex responses presented below are due solely to the imposition of $\beta$. Thus, in all experiments we first simulate a 5 day spin-up period with $\beta=0$, after which $\beta$ is instantaneously turned on to a constant value for the subsequent 95 days or shorter if the vortex has reached quasi-equilibrium. Also, since a vortex on a $\beta$-plane will gradually drift with time, we define the centroid of the vorticity to track the vortex center with time and use this to calculate a storm-centered radial profile of the tangential wind at each time-step.
\begin{figure*}[t]
 \centerline{\includegraphics[width=38pc]{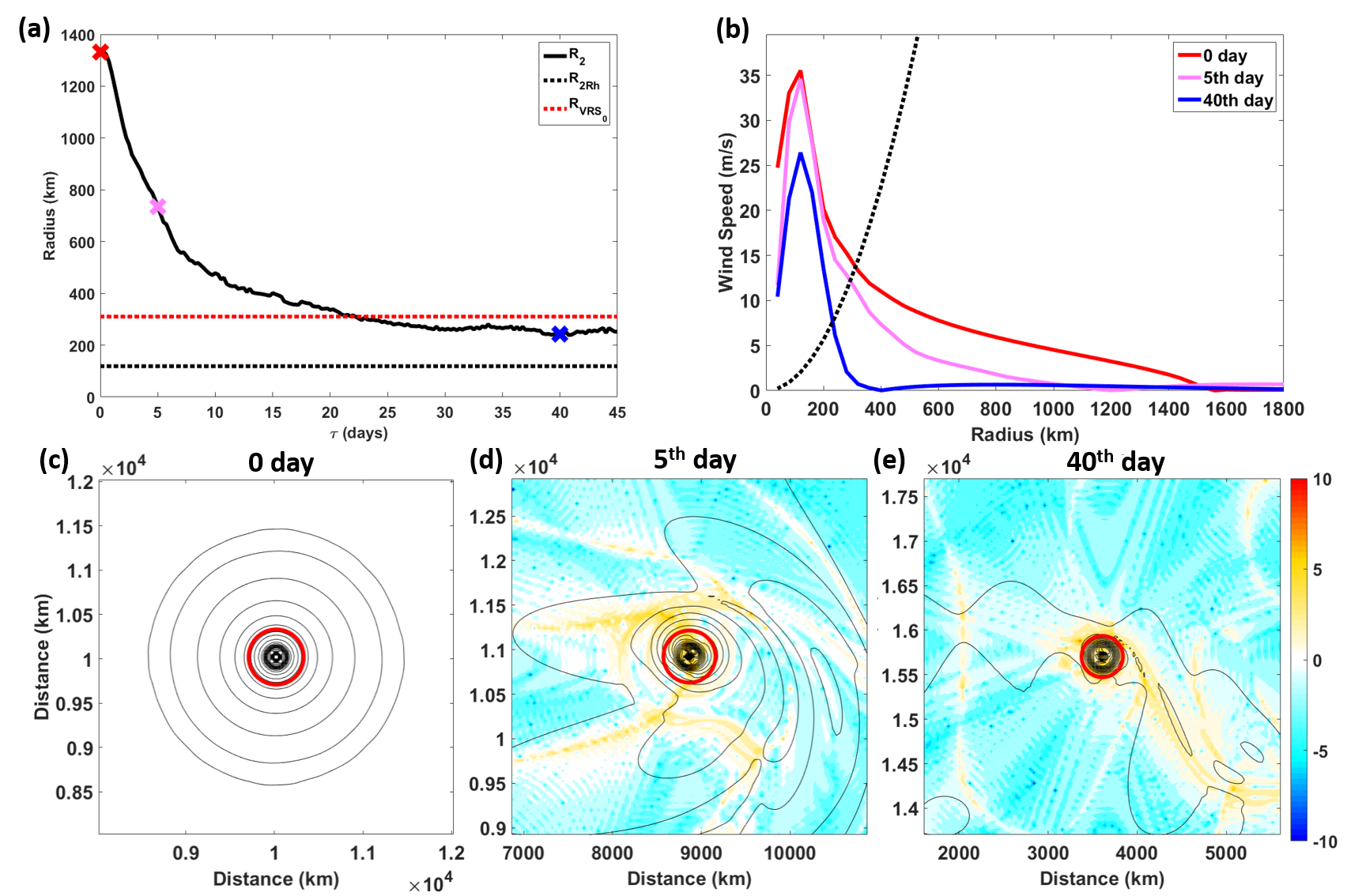}}
 \caption{Results of $CTRL$. (a) $R_2$ time series, where $\tau$ indicates time since $\beta$ is turned on; markers indicate times from the initial stage (day 0, red cross), the shrinking stage (day 5, pink cross), and the quasi-equilibrium stage (day 40, blue cross), and dashed line represents the Rhines scale of $2ms^{-1}$ wind ($R_{2Rh}$, eq \ref{eq:RhinesScale}). (b) $U_c$ profiles at each of the three stages. (c)-(e) Wind speed (black contours), $log_{10}Rh$ (shading), and $R_{VRS}$ (red circle) at each of the three stages, respectively.}
  \label{fig:CTRL}
\end{figure*}
\subsubsection{Experiment set "$VARYVMAX$": Varying vortex initial intensity}
In our experiments, resolution limitations will cause vortices with different initial sizes to have different initial intensities because the inner-core is poorly resolved; this effect acts similar to a radial mixing and hence acts to decrease $V_{max}$ and increase $R_{max}$. 
Thus, we design experiment set "$VARYVMAX$" (Figure \ref{fig:Initial_Condition}a), which has vortices with different initial intensities at fixed size to examine the impact of vortex intensity on the evolution of the vortex circulation. Note that the outer tangential wind structure remains constant as intensity is varied, which is by design in the C15 model as the inner and outer circulations of TCs tend to vary independently. 

\subsubsection{Experiment "$VARYR0$": Varying vortex initial size} 
To investigate the effect of the Rhines scale on vortices with different sizes, we design experiment set "$VARYR0$" (Figure \ref{fig:Initial_Condition}b), which has initial wind profiles specified using $R_0$ over a range of values from 600 to 1800 km (see Table \ref{t1}) with $\beta$ fixed at the $CTRL$ value. Since all the members in this experiment have exact same value of $\beta$, they all have the exactly same $U_{Rh}$ profile. Thus, when increasing $R_0$, the vortex wind profile expands outward, and as a result ${R}_{VRS}$ increases as $R_0$ increases.

\subsubsection{Experiment set $VARYBETA$: Varying $\beta$}
To investigate the effect of the Rhines scale on vortex size at different values of $\beta$, we design experiment "$VARYBETA$" (Figure \ref{fig:Initial_Condition}c), which imposes the CTRL wind profile for all members on a $\beta$-plane with $\beta$ over a range of values from $2.2547 \times 10^{-11}$ $m^{-1}s^{-1}$ to $3.9756 \times 10^{-12}$ $m^{-1}s^{-1}$, corresponding to a latitude of $10^{\circ}$ N to $80^{\circ}$ N on Earth. As noted above, $f$ in the C15 model is held constant at its CTRL value to isolate effects of $\beta$ on vortex evolution at fixed initial vortex structure. The Rhines speed increases at all radii with increasing $\beta$ (Eq. \ref{eq:RhinesSpeed}), and as a result ${R}_{VRS}$ decreases as $\beta$ increases. 

\begin{figure*}[t]
 \centerline{\includegraphics[width=38pc]{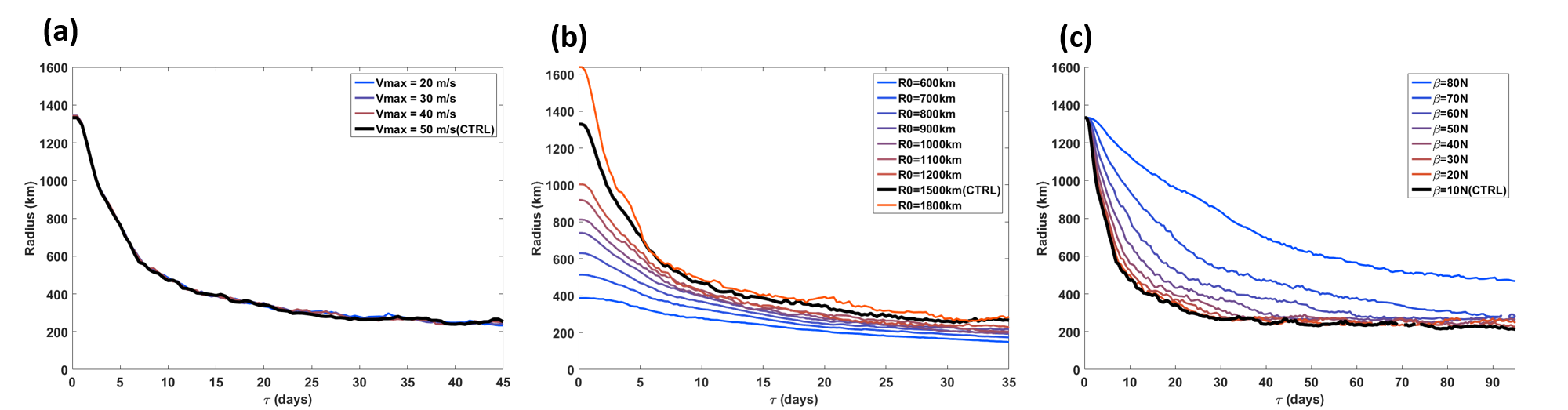}}
 \caption{$R_2$ time series for all three experiment sets. (a) $VARYVMAX$. (b) $VARYR0$. (c) $VARYBETA$. $CTRL$ is highlighted in thicker black curves across experiment sets.
 }
 \label{fig:R2_Timeseries}
\end{figure*}

\section{Results}

\subsection{Vortex response: CTRL}

We begin by describing the simulated response of our Control vortex to $\beta$ and place it in the context of the vortex Rhines scale. 
Figure \ref{fig:CTRL} shows the structural evolution of $CTRL$.
Figure \ref{fig:CTRL}a shows the time series of the radius of $2$ $ms^{-1}$ wind (hereafter $R_2$) of this vortex, which we use as our measure of the overall size of the storm, as well as the value of the Rhines scale evaluated at $2 ms^{-1}$ given by $R_{2Rh} = R_{Rh}(U_c=2ms^{-1})$ (Eq.\ref{eq:RhinesScale}), and the vortex Rhines scale ($R_{VRS_0}$) of this vortex; $R_2$ shrinks toward a quasi-equilibrium value close to $R_{VRS_0}$. Note that the equilibrium value of $R_2$ is still larger than $R_{2Rh}$. However, the relationship between final equilibrium size and the Rhines scale magnitude depends strongly on the choice of wind speed used to define size; in the limit of $U_c=0ms^{-1}$, the Rhines scale is zero. This demonstrates a shortcoming of the traditional Rhines scale for defining a precise limit on vortex size, which motivates the use of the vortex Rhines scale, which does not require choosing an arbitrary wind speed, in our subsequent analyses.

Figure \ref{fig:CTRL}b-e displays a detailed 2D analysis of the vortex at different stages of its evolution based on Figure \ref{fig:CTRL}a: the initial state (day 0), the shrinking stage (day 5), and the quasi-equilibrium stage (day 40). Figure \ref{fig:CTRL}b shows the azimuthal-mean tangential wind profile at each stage. In order to demonstrate the dominant term in Eq.\ref{eq:BVE}, we calculate the base-10 logarithm of the Rhines number, $\log_{10}Rh$, at each grid point (a value of zero corresponds to $Rh = 1$). Figure \ref{fig:CTRL}c-e show storm-centered maps of absolute wind speed and $\log_{10}Rh$ at each stage, with warmer colors (positive $\log_{10}Rh$) representing the vortex-dominant regime and cooler colors (negative $\log_{10}Rh$) representing the wave-dominant regime and with $R_{VRS}$ shown as a red circle.
Initially, the vortex's $R_2$ is larger than the $R_{2Rh}$ but no asymmetric structure has yet developed within the vortex. During the shrinking stage, the vortex size decreases rapidly, as the outer circulation outside of $R_{VRS}$ is highly variable and azimuthally asymmetric compared to inner core circulation inside of $R_{VRS}$.  At small radii, the nonlinear term is generally dominant, which is evidence by the warmer colors (positive $\log_{10}Rh$); at larger radii the wave term is generally dominant, which is evidence by the colder colors (negative $\log_{10}Rh$). Note that $R_{VRS}$ approximately separates the two regions. Finally, in the quasi-equilibrium stage, the circulation has nearly vanished in the wave region outside of $R_{VRS}$ while it remains intact and highly axisymmetrized inside of $R_{VRS}$.

$CTRL$ demonstrates how the radial structure of the response of a single vortex on a $\beta$-plane can be described at least qualitatively via $R_{VRS}$. The vortex Rhines scale appears to impose a strong limit on vortex size by dividing the vortex into two regions with distinct dynamical characteristics. Circulations in the vortex region produce minimal Rossby waves and instead are simply self-advected, thus maintaining a highly axisymmetric structure. In contrast, circulations in the wave region generate significant Rossby wave activity that produce a highly azimuthally-asymmetric structure that acts to spin down the circulation there.  

\subsection{Response with varying $V_{max}$}

Since we have intensity variability within our experiment members due to resolution limitations, it is important to demonstrate that changes in inner core intensity do not affect the outer circulation before we analyze any experiments systematically.
Figure \ref{fig:R2_Timeseries}a shows the $R_2$ time series of all members in $VARYVMAX$.
All members exhibit a nearly identical size evolution across experiments. These results indicate that variations in intensity changes the wind speeds in the vortex region ($R < R_{VRS}$) but not the broad outer circulation. Note that in a barotropic model there is no secondary circulation that links the intensity change in inner core to the distant outer circulation, as there would be in a real TC. However, the low-level circulation of a TC is characterized by inflow at nearly all radii and so the outer circulation would not be expected to directly feel changes in inner-core structure, a behavior also common to observed and simulated TCs \citep{Frank_1977,Merrill_1984,Chavas_and_Lin_2016,Rotunno_and_Bryan_2012}.
\begin{figure*}[t]
 \centerline{\includegraphics[width=38pc]{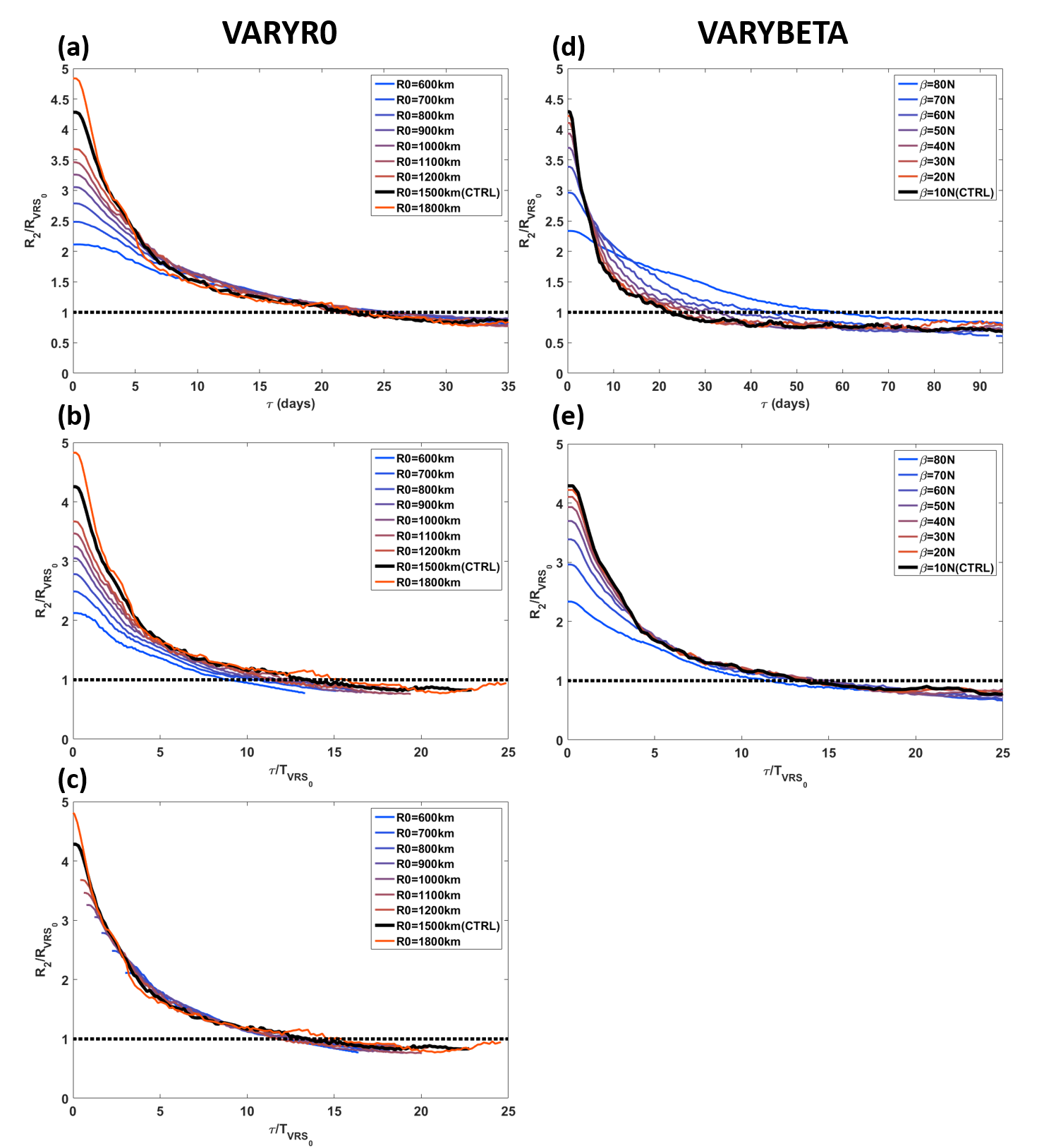}}
 \caption{Results from Figure \ref{fig:R2_Timeseries}b-c for VARYR0 (left column) and VARYBETA in non-dimensional space (right column). (a) $R_2/R_{VRS_0}$ vs $\tau$ for $VARYR0$. (b) Same as (a) but with time non-dimensionalized  by $T_{VRS_0}$, $\tau/T_{VRS_0}$. (c) Same as (b), but with curves are shifted in time to align with the $CTRL$. (d) Same as (a) but for $VARYBETA$.  (e) Same as (b) but for $VARYBETA$.
 }
 \label{fig:EXP_Normalized}
\end{figure*}

\subsection{Responses with varying initial $R_0$ or $\beta$}

Next, Figure \ref{fig:R2_Timeseries}b shows the $R_2$ time series of all members in $VARYR0$. Warmer colors indicate members with larger initial vortex size. 
In $VARYR0$, vortices with larger initial size shrink in size faster, but they all gradually converge in size and eventually reach a quasi-equilibrium of approximately 200-300 km.

Figure \ref{fig:R2_Timeseries}c shows the $R_2$ time series of all members in $VARYBETA$. Warmer colors indicate members with larger $\beta$ (lower Earth latitude). Each vortex in $VARYBETA$ has exactly the same $U_{c}$ profile and hence all time series start from the same $R_2$ value. In $VARYBETA$, experiments at larger $\beta$ (lower latitude) will shrink in size faster, 
while most vortices reach quasi-equilibrium with broadly similar $R_2$.

The $R_2$ time series analysis demonstrates the systematic behavior and their differences between $VARYR0$ and $VARYBETA$. All members in both experiments shrink in size with different rates after $\beta$ is turned on.
Vortex size shrinks at a faster rate for larger initial $R_0$ at fixed $\beta$ or for larger $\beta$ at fixed initial $R_0$.

This similarity arises because all members from each experiment have different initial sizes relative to their vortex Rhines scale. Therefore we next examine the size evolution of each member in a non-dimensional sense relative to their $R_{VRS}$ to provide more general physical insight into the results of these experiments.

\subsection{Vortex size evolution: Non-dimensional space}
To generalize the vortex size evolution across $VARYR0$ and $VARYBETA$, we next examine the evolution of $R_2$ non-dimensionalized by the initial $R_{VRS}$ ($R_{VRS_0}$), ${R_2}/{R_{VRS_0}}$, for each member.

We begin with $VARYR0$. Figure \ref{fig:EXP_Normalized}a shows ${R_2}/{R_{VRS_0}}$ for $VARYR0$. Warmer colors represent members with larger initial ${R_2}/{R_{VRS_0}}$. As mentioned above, each member in $VARYR0$ has different initial ${R_2}/{R_{VRS_0}}$ due principally to the different initial vortex size. Experiments with higher initial ${R_2}/{R_{VRS_0}}$ will decrease in size faster, especially for the first 5 days of the experiment, which is a strong evidence in support of our hypothesis. All members' ${R_2}/{R_{VRS_0}}$ eventually converge to a value slightly smaller than $1$ in quasi-equilibrium stage, indicating that each vortex shrinks to a size slightly smaller than its initial $R_{VRS}$. Moreover, all experiments' time-series converge to nearly the same curve as they approach the quasi-equilibrium stage, in contrast to the dimensional case (Fig \ref{fig:R2_Timeseries}). This result indicates that $R_{VRS_0}$ imposes a strong limit on equilibrium vortex size. Note that, though each experiment starts from a different initial ${R_2}/{R_{VRS_0}}$, they all reach their quasi-equilibrium stage at similar times, indicating that they have different shrinking rates but a similar overall equilibration timescale. 

We next propose to non-dimensionalize experiment time $\tau$ by the initial vortex Rhines timescale, $T_{VRS_0}$ (Eq. \ref{eq:TVRS}). To test whether $T_{VRS_0}$ can represent the vortex size shrinking time scale across experiments, we non-dimensionalize time with the value of $T_{VRS_0}$ to test if the curves will further collapse together. Note a time-varying $T_{VRS}$ can't be used here; a single constant time-scale must be chosen. Figure \ref{fig:EXP_Normalized}b shows the evolution of ${R_2}/{R_{VRS_0}}$ vs $\tau/{T_{VRS_0}}$ for $VARYR0$, which produces a similar shrinking rate during the shrinking stage across experiments but separates the curves thereafter. Hence, as a final step, we shift them in time to align with the curve with the $CTRL$ (see Figure \ref{fig:EXP_Normalized}c). The final result yields curves that approximately collapse to a single universal curve. This outcome indicates that $T_{VRS_0}$ represents the timescale associated with the shrinking rate of each experiment, and that the evolution depends only on the current value of ${R_2}/{R_{VRS_0}}$.

Next we analyze $VARYBETA$. Figure \ref{fig:EXP_Normalized}d is the same as Figure \ref{fig:EXP_Normalized}a but for $VARYBETA$. Each member has different initial ${R_2}/{R_{VRS_0}}$ due to the different initial $R_{VRS}$. Similar to $VARYR0$, members in $VARYBETA$ with larger initial ${R_2}/{R_{VRS_0}}$ also shrink faster and also converge to a value smaller than $1$ in the quasi-equilibrium stage. However, in contrast to $VARYR0$, members in $VARYBETA$ equilibrate over significantly different timescales, especially for members with lower $\beta$, indicating that they have different underlying timescales. Thus, we non-dimensionalize time in Figure \ref{fig:EXP_Normalized}e, which shows ${R_2}/{R_{VRS_0}}$ vs $\tau/{T_{VRS_0}}$. Beyond an initial period of rapid shrinking (i.e. non-dimensional time equals to 5 onwards), the simulations now collapse well through to equilibrium, such that all members reach equilibrium at the same time. The curves do not collapse closely during shrinking stage, indicating perhaps some additional dynamics at play that cannot be captured via our two dominant vortex Rhines scale parameters; note that a time translation similar to that done for $VARYR0$ will not help further collapse these curves since they shrink at different rates. 

\begin{figure}[t]
 \centerline{\includegraphics[width=19pc]{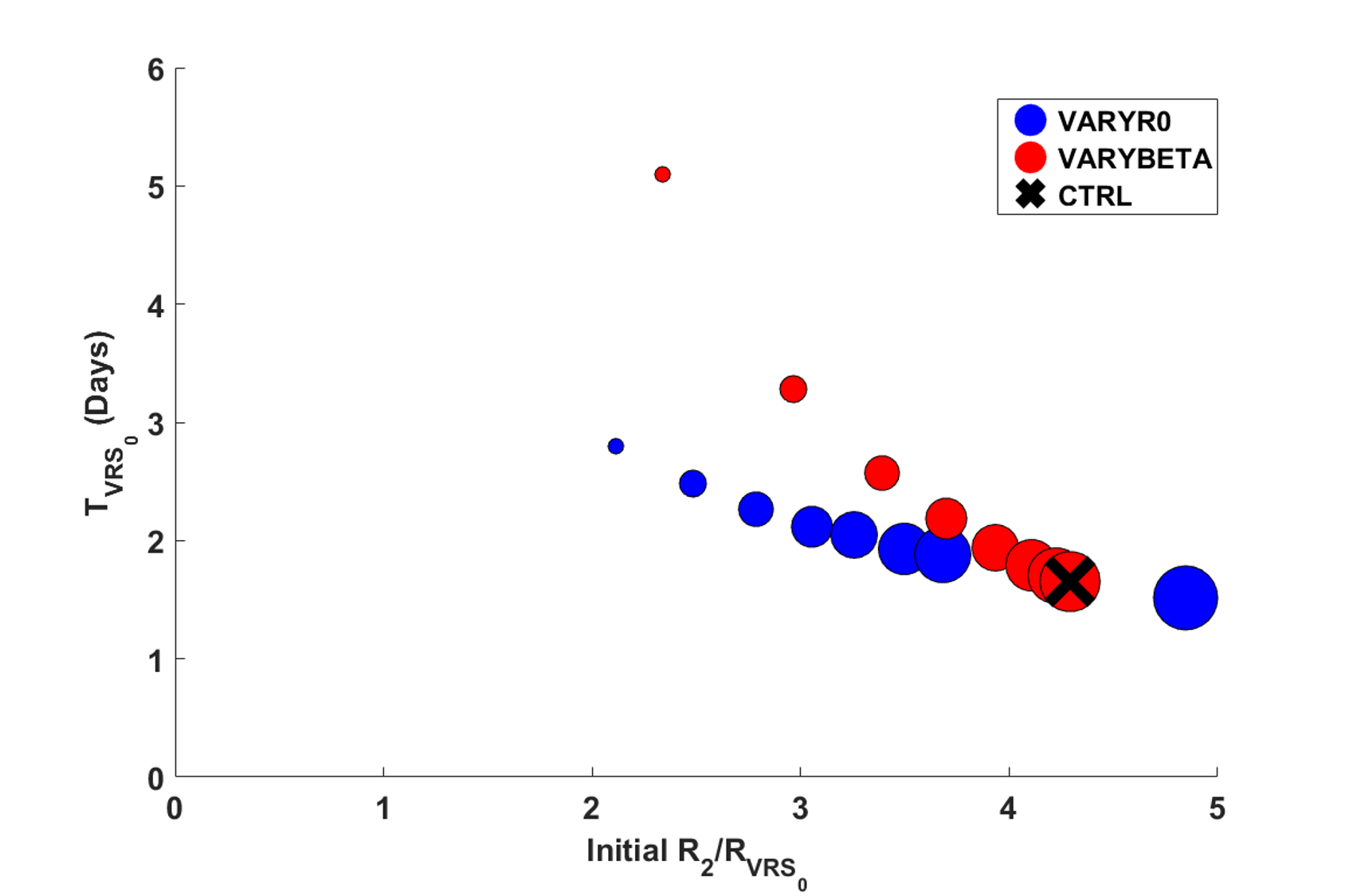}}
 \caption{Scatter plot of $T_{VRS_0}$ against initial $R_2/R_{VRS_0}$ for experiment sets VARYR0 and VARYBETA. $CTRL$ highlighted with a black cross. Larger marker size represents member with larger magnitude $R_0$ or $\beta$ within the relevant experiment set.}
 \label{fig:EXP_TVRS}
\end{figure}

To understand how $T_{VRS_0}$  varies across experiments, Figure \ref{fig:EXP_TVRS} shows the relation between $T_{VRS_0}$ and the initial $R_2$/$R_{VRS_0}$ of all members in $VARYR0$ and $VARYBETA$. In $VARYBETA$, a vortex at larger $\beta$ (lower latitude) at fixed initial $R_0$ will have a smaller $R_{VRS_0}$ but a larger $U_{VRS_0}$, which both act to decrease $T_{VRS_0}$ (Eq. \ref{eq:TVRS}); thus, $T_{VRS_0}$ decreases rapidly as $\beta$ is increased. On the other hand, in $VARYR0$, a vortex with a larger initial $R_0$ at fixed $\beta$ will have a larger $R_{VRS_0}$ and $U_{VRS_0}$, whose effects on $T_{VRS_0}$ oppose one another; thus,  $T_{VRS_0}$ decreases slowly as $R_0$ is increased. Thus, the behavior of $T_{VRS,0}$ differs when $R_2$/$R_{VRS}$ is varied by changing $R_0$ vs $\beta$. Basic physical insight into the meaning of this timescale is provided in subsection \ref{section_Trw} below.
Across both experiment sets, our results show that non-dimensionalization of radius and time by $R_{VRS_0}$ and  $T_{VRS_0}$, respectively, can allow all curves to significantly collapse with each other. The broad implication is that $R_2$ shrinks toward $R_{VRS_0}$ over a fundamental timescale given by $T_{VRS_0}$, though it occurs in slightly different ways when varying $R_0$ vs. $\beta$. 

\begin{figure*}[t]
 \centerline{\includegraphics[width=38pc]{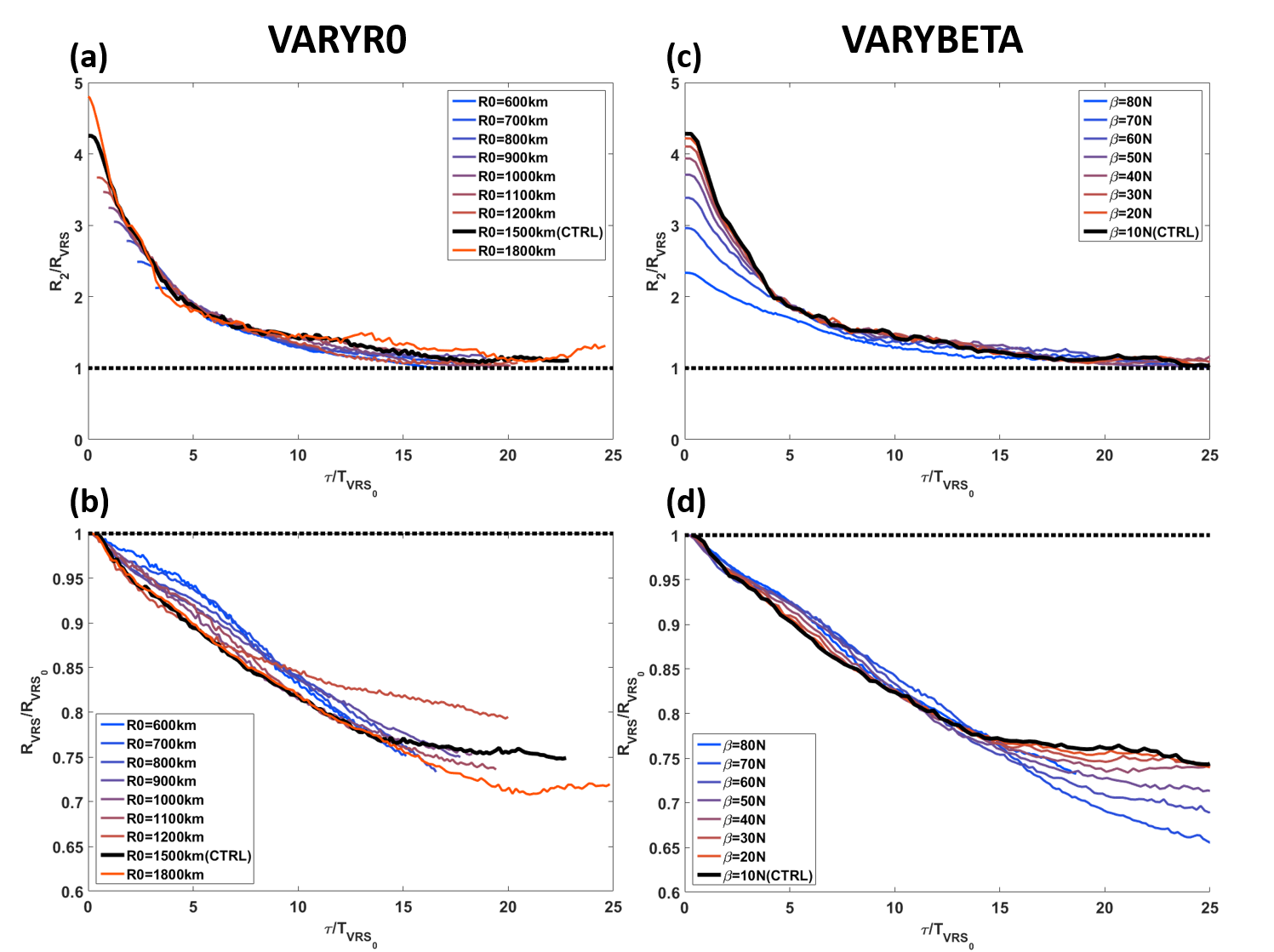}}
 \caption{Non-dimensional evolution of $R_2$ from $R_{VRS}$, and $R_{VRS}$ from its initial value $R_{VRS_0}$. (a) $R_2/R_{VRS}$ vs $\tau/T_{VRS_0}$ for $VARYR0$, and curves are shifted in time to align with the $CTRL$. (b) $R_{VRS}/R_{VRS_0}$ vs $\tau/T_{VRS_0}$ for $VARYR0$. (c) Same as (a) but for $VARYBETA$, and without time translation. (d) Same as (b) but for $VARYBETA$}
 \label{fig:EXP_RVRSRVRS0}
\end{figure*}

For $VARYR0$, the curves collapse at all times via a time translation, indicating that there is a universal non-dimensional shrinking rate that only depends on the current value of $R_2/R_{VRS_0}$ (i.e. the size evolution is path-independent), and $T_{VRS_0}$ represents the timescale of this rate. For $VARYBETA$, each curve is initially different but all curves converge to 1 over a specific single timescale, and $T_{VRS_0}$ represents this overall equilibration timescale. Why this distinction arises between the two experiment types is not currently known but may be related to the wave dynamics in the outer region that differs when varying storm size ($R_0$ of $U_c$ profile) vs. varying $\beta$ (slope of $U_{Rh}$).

\begin{figure*}[t]
 \centerline{\includegraphics[width=38pc]{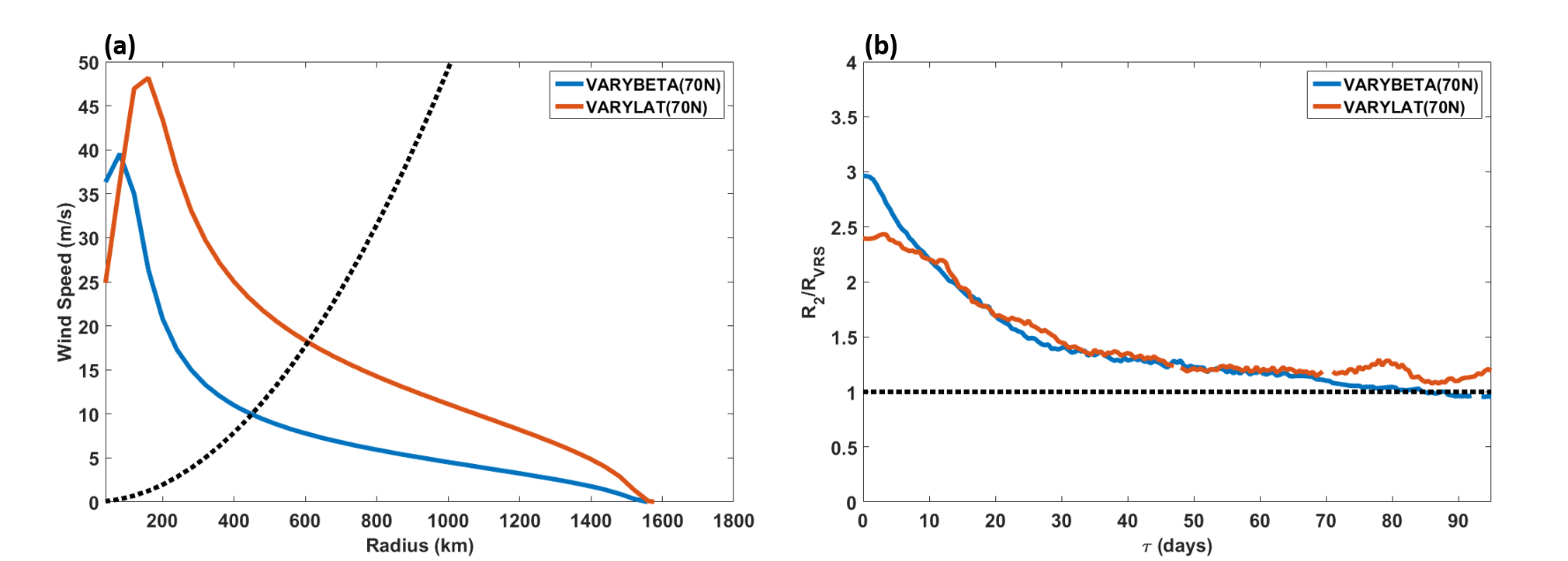}}
 \caption{Comparison of $70^{\circ}$N experiment from $VARYBETA$ with $VARYLAT(70N)$. (a) Initial $U_c$ profiles and their $U_{Rh}$ profile (dashed curve). (b) $R_2/R_{VRS}$.}
 \label{fig:Vary_Lat}
\end{figure*}

\subsection{The evolution of the vortex Rhines scale}

We have demonstrated how we can understand the vortex size evolution using \textit{initial} values of $R_{VRS_0}$ and $T_{VRS_0}$. 
We found that knowledge of the initial vortex structure and $\beta$ value alone are sufficient to predict both the upper limit on equilibrium vortex size ($R_{VRS_0}$) and the shrinking rate ($T_{VRS_0}$). However, $R_{VRS}$ may also change with time during a given experiment.
We may also use the time-dependent $R_{VRS}$ to normalize $R_2$ (Figure \ref{fig:EXP_RVRSRVRS0}a and c), which yields very similar results to that presented above using $R_{VRS_0}$: the lone difference is that the $R_2/R_{VRS}$ curves converge to an equilibrium value that is almost exactly 1, rather than a bit smaller than 1 in Fig \ref{fig:EXP_Normalized}, indicating that the final equilibrium size is exactly given by $R_{VRS}$. Although normalizing by the time-dependent $R_{VRS}$ is technically more precise, it requires knowledge of the vortex evolution itself and hence is no longer a true prediction. 

$R_{VRS_0}$ may be used instead of the time-dependent $R_{VRS}$ because the intrinsic timescale of $R_{VRS}$ relative to its initial value $R_{VRS_0}$ also follows $T_{VRS_0}$. (Recall that $T_{VRS_0}$ must be used, as a time-varying $T_VRS$ does not make sense for our analysis.) The relation between $R_2/R_{VRS}$ and $R_2/R_{VRS_0}$ can be written as following:
\begin{equation}\label{eq:RVRS0}
\frac{R_2}{R_{VRS_0}} = \frac{R_{VRS}}{R_{VRS_0}}\frac{R_2}{R_{VRS}}.
\end{equation}
Mathematically, if the evolution of $R_2/R_{VRS}$ and $R_{VRS}/R_{VRS_0}$ both collapse across experiments after non-dimensinoalizing by $T_{VRS_0}$, then $R_2/R_{VRS_0}$ will collapse as well. Figure \ref{fig:EXP_RVRSRVRS0}b and d shows $R_{VRS}/R_{VRS_0}$ vs $\tau_{VRS}/T_{VRS_0}$ for $VARYR0$ and $VARYBETA$, respectively. For both experiment sets, all curves nearly collapse, indicating that $R_{VRS}$ decreases relative to their initial values at the same non-dimensional rate.

For this reason, $R_2/R_{VRS}$ and $R_2/R_{VRS_0}$ yield similar non-dimensional results. Using the initial value is therefore preferable since as it has already be known in the first place, it also can provide the prediction of how vortex size on a $\beta$-plane will evolve with time. This result further highlights how $T_{VRS_0}$ is the dominant intrinsic timescale for this system.

\subsection{A more Earth-like case: allowing $f$ to vary}
In VARYBETA we only modified the $\beta$-plane value in the barotropic model while leaving $f$ constant when specifying the initial vortex using the C15 model. In the real world on a rotating sphere, changing latitude will change $\beta$ and $f$ simultaneously. Allowing $f$ to change in the C15 model while holding $R_0$ fixed will change the radial structure of the resulting wind profile inside of $R_0$.

Here we briefly test whether changing $f$ consistent with a change in $\beta$ (i.e. true changes in latitude) on Earth will affect our result significantly. To do so, we perform an experiment identical to our $VARYBETA$ experiment at 70N except with a vortex generated by the C15 model with $f$ set at its value at 70N (hereafter $VARYLAT(70N)$), and compare the results. 
Figure \ref{fig:Vary_Lat}a shows the initial $U_{c}$ profiles; the $U_{Rh}$ profile is identical for each.
Larger $f$ shifts the wind field structure radially outwards towards $R_0$, including a larger $R_{max}$ (which results in larger $V_{max}$ when inserted into the barotropic model) and stronger winds at most radii beyond the inner core. 

Figure \ref{fig:Vary_Lat}b shows $R_2$/$R_{VRS}$ between these two experiments. Results show that despite different initial $R_2$/$R_{VRS}$, the two members from $VARYBETA$ and $VARYLAT(70N)$ have similar $R_2$/$R_{VRS}$ evolution after 10 days. This indicates that our overall results are not dependent on holding $f$ fixed in the wind profile and hence may be directly applicable to the real Earth.

\subsection{Linking $T_{VRS}$ to the Rossby wave group velocity} \label{section_Trw}

The dynamical details of the vortex-wave interaction, in particular a mechanistic understanding of the waves themselves and the energy transfer that they induce, are not tackled in this work. Here, though, we provide a simple first step in this direction by linking $T_{VRS}$ to the Rossby wave group velocity. We defined $T_{VRS}$ as the overturning timescale of the circulation at $R_{VRS}$, which can be written as a function of $R_{VRS}$ and $\beta$:
\begin{equation}\label{eq:TVRS_RVRS}
T_{VRS}=\frac{2{\pi}R_{VRS}}{U_{VRS}}=\frac{2{\pi}R_{VRS}}{2\pi\beta{R_{VRS}}^2}=\frac{1}{\beta{R_{VRS}}}.
\end{equation}
where we have made use of the fact that the $U_{VRS}$ can be defined using the definition of the Rhines speed (Eq. \ref{eq:RhinesSpeed}) as $U_{Rh}(r=R_{VRS})$. Following the theory discussed above, at $R_{VRS}$ the circulation's overturning timescale should equal the planetary Rossby wave generation timescale, which we propose should be directly related to the timescale of planetary Rossby wave propagation, $T_{RW}$. Here, similar to how we calculate $T_{VRS}$, we estimate the $T_{RW}$ by using circulation's circumference as length scale, and the group velocity of the planetary Rossby wave ($c_{gRW}$) as its propagation speed:
\begin{equation}\label{eq:TRW_Ori}
T_{RW}\propto\frac{2{\pi}R_{VRS}}{c_{gRW}}.
\end{equation}
For a barotropic planetary Rossby wave, its group velocity is as following \citep[][p.228]{Vallis_2017_p.446}:
\begin{equation}\label{eq:CgRW_Ori}
c_{gRW}=\sqrt{{c_g^x}^2+{c_g^y}^2}=\sqrt{\left(\frac{\beta(k^2-l^2)}{(k^2+l^2)^2}\right)^2+\left(\frac{2kl\beta}{(k^2+l^2)^2}\right)^2},
\end{equation}
where $k$ and $l$ are the wavenumbers in the $x$ and $y$ direction, respectively. $c_g^x$ and $c_g^y$ are the Rossby wave group velocity in the $x$ and $y$ direction, respectively. For a vortex we assume axisymmetry, such that $k=l$, which results in zero group velocity in the $x$ direction. We take wavenumber to be inverse proportional to the circulation's circumference ($k=1/(2 \pi R_{VRS})$). The group velocity at $R_{VRS}$ may then be written as:
\begin{equation}\label{eq:CgRW_RVRS}
c_{gRW}=c_g^y=\frac{2kl\beta}{(k^2+l^2)^2}=\frac{2k^2\beta}{(k^2+k^2)^2}=\frac{\beta}{2k^2}=2\beta \pi^2R_{VRS}^2.
\end{equation}
And now we can substitute Eq.\ref{eq:CgRW_RVRS} into Eq.\ref{eq:TRW_Ori} to write $T_{RW}$ in terms of $R_{VRS}$:
\begin{equation}\label{eq:TRW_RVRS}
T_{RW}\propto\frac{2{\pi}R_{VRS}}{2\beta \pi^2R_{VRS}^2}=\frac{1}{\pi\beta {R_{VRS}}}.
\end{equation}
Comparing Eq.\ref{eq:TRW_RVRS} to Eq.\ref{eq:TVRS_RVRS}, we find that $T_{RW}$ is identical in form to $T_{VRS}$, differing only by a factor $\pi$. The similarity between $T_{RW}$ and $T_{VRS}$ indicates that $T_{VRS}$ is proportional to the planetary Rossby waves propagation timescale at $R_{VRS}$, which should be directly related to the wave-induced dissipation of the vortex outer circulation that drives the size evolution of a TC-like vortex on a $\beta$-plane.
This basic theoretical linkage, in conjunction with our results and conceptual understanding above, provides further insight into how the vortex Rhines scale governs the first-order dynamics of the vortex response to $\beta$. Understanding the detailed radial structure of these waves may provide deeper insight and is left for future work.

\begin{figure*}[t]
 \centerline{\includegraphics[width=38pc]{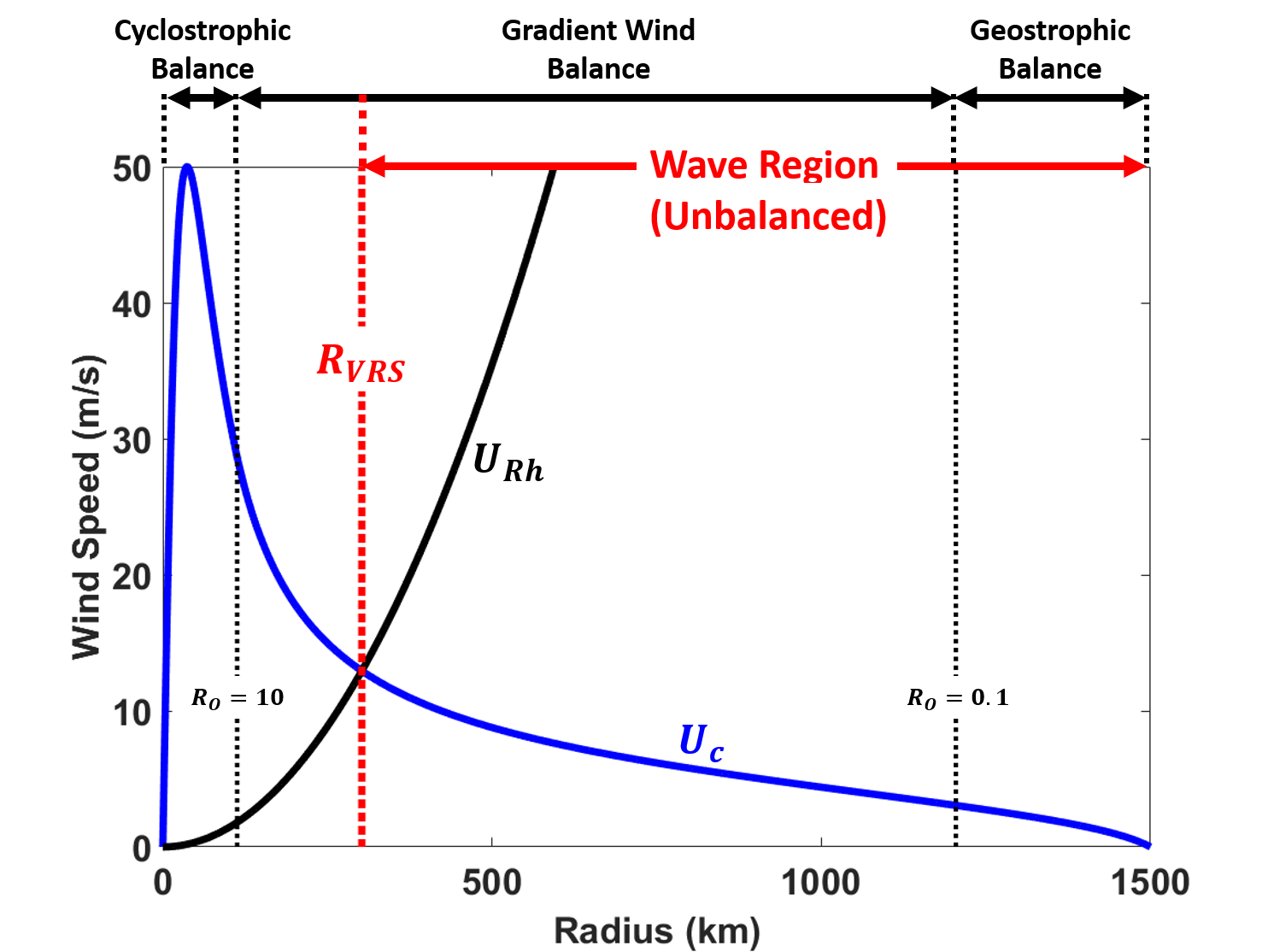}}
 \caption{Conceptual diagram illustrating the different dynamical regimes in the $CTRL$ (10N) initial vortex wind profile (blue), with Rhines speed profile (black). The dashed black vertical lines indicate radii of different Rossby numbers ($R_o$) (10, 1, and 0.1 from small to large radii), and the red dashed vertical line represents the $R_{VRS}$. Neglecting $\beta$, the region with $R_0>10$ corresponds to cyclostrophic balance, $10>R_0>1$ gradient wind balance, and $R_0<1$ geostrophic balance, respectively. $\beta$ introduces an unbalanced wave-generation region beyond $R_{VRS}$.}
 \label{fig:Conceptual_RossbyNo}
\end{figure*}

\section{Conclusion and discussion}

This study derives a new concept called the vortex Rhines scale and applies it to experiments with a barotropic model to understand how $\beta$ limits the size of a tropical cyclone-like vortex.
Since the barotropic model governing equation includes only the advection of the relative and planetary vorticity, our experiment design provides an idealized and straightforward framework to investigate the dynamics of a TC-like vortex in the presence of $\beta$ and its relationship to the traditional Rhines scale.

The key findings of this study are as follows:
\begin{enumerate}
  \item We derive a quantity called the vortex Rhines scale ($R_{VRS}$), which translates the traditional Rhines scale into the context of an individual axisymmetric vortex, and show how it can be used to understand the effect of $\beta$ on a TC-like vortex. 
  \item The vortex Rhines scale serves as a robust limit on the size of a TC-like vortex on a barotropic $\beta$-plane, which corroborates the finding in \citet{Chavas_and_Reed_2019} that storm size scales with the traditional Rhines scale. The circulation beyond the vortex Rhines scale will weaken with time, which manifests itself as a shrinking of vortex size. $R_{VRS}$ offers a more useful scale for the limit of TC size than the traditional Rhines scale.
  \item Theoretically, the vortex will be divided into two regions by $R_{VRS}$: vortex region at smaller radii and wave region at larger radii. In the vortex region, the circulation is highly axisymmetric and largely unaffected by $\beta$. In the wave region, planetary Rossby wave generation is strong and waves distort and dissipate the outer circulation.
  \item A larger vortex relative to its $R_{VRS}$ will shrink faster, and all vortices shrink towards an equilibrium close to its vortex Rhines scale. 
  \item Vortex size shrinks toward $R_{VRS}$ following a dominant timescale given by $T_{VRS}$, though the role of that timescale differs slightly when varying $R_0$ vs $\beta$. $T_{VRS}$ is also shown to be closely related to the Rossby wave group velocity at the vortex Rhines scale, which provides a direct link between our theory and the dynamics of the waves themselves.
  \item The first-order evolution of the vortex for any value of $R_0$ and $\beta$ is controlled by the initial value of $R_{VRS}$ and $T_{VRS}$, thereby enabling one to predict the vortex response from the initial condition alone. 
  \item A similar outcome occurs when allowing $f$ in the initial vortex structure to change consistent with $\beta$, indicating that the results are also applicable to an Earth-like setting.
  
\end{enumerate}

It is important to place our idealized results in the context of real TCs on Earth. The typical duration of a TC is on the order of 10 days \citep{Webster_et_al_2005} and form from pre-existing disturbances that may have propagated for much longer. Since most of our experiment members (especially for those on a lower latitude $\beta$-plane) have shrunk rapidly within 10 days, the vortex Rhines scale would be expected to strongly limit TC size on Earth and hence explain why storm size appears to follow a Rhines-type scaling in observations and models. Moreover, it explains how storm size may vary widely in nature, as TC size may remain steady at any size that is reasonably small relative to this scale, since the vortex Rhines scale only sets an upper bound on vortex size.
Additionally, $R_{VRS}$ in our experiments range from $200$ to $400$ km, which is substantially smaller than the $f^{-1}$ theoretical length scale for TC size on the $f$-plane, $V_p/f$, where $V_p$ is the potential intensity \citep{Chavas_and_Emanuel_2014}, which is larger than $1000$ km at low latitudes (and goes to infinity at the equator). In other words, if a TC were for form with a size equal to $V_p/f$, our results indicate that $\beta$ and its induced wave effects would cause it to shrink rapidly within a few days. This is a simple mechanistic explanation for why $V_p/f$ is not an appropriate scaling for TC size in the tropics as has been noted in observations \citep{Chavas_et_al_2016}. 
Finally, though our derivations and analyses have assumed axisymmetry, these physics may be general to any vortex, even asymmetric ones such as extra-tropical cyclones. Indeed, our results and theory suggests a simple mechanistic explanation for why the Rhines scale cuts of the upscale energy cascade in 2D turbulence and limits extra-tropical cyclone size \citep[e.g.,][]{Held_and_Larichev_1996,Chai_and_Vallis_2014,Chemke_and_Kaspi_2015,Chemke_and_Kaspi_2016,Chemke_et_al_2016}.

All of our experiments are barotropic. This simplified approach omits various key physical features in real TCs, but it is unclear how non-barotropic effects would modify these barotropic responses. For example, radial inflow in TC will communicate the reduction in angular momentum in the outer region to the inner core, so the inner core wind field would be expected to shrink too.
Further investigation in models with higher degrees of complexity and in observations is needed to evaluate our findings in a full-physics baroclinic environment. 

Finally, to provide a broader perspective on our findings, we conclude with a simple conceptual diagram to place the effect of $\beta$ in the context of standard dynamical balance regimes common in atmospheric science. Figure \ref{fig:Conceptual_RossbyNo} illustrates different dynamical regimes as a function of radius for an example TC wind profile. On an $f$-plane ($\beta = 0$), a vortex wind profile can remain steady at all radii and its size remain unchanged. This is because on the $f$-plane at any given radius the flow can exist in a balanced state, where the specific balance is commonly defined using the Rossby number. Within the inner core region (inside of approximately 100 km in Figure \ref{fig:Conceptual_RossbyNo}), the Rossby number of the circulation is significantly larger than 10, which corresponds to cyclostrophic balance. Between the inner core and far outer circulation, the Rossby number ranges between 0.1 and 10, which corresponds to gradient wind balance. Finally, in the far outer circulation inside of the outer radius, the Rossby number is less than 0.1, which corresponds to geostrophic balance. However, in the presence of non-zero $\beta$, there is now an unbalanced regime at radii beyond the vortex Rhines scale, which corresponds to the wave region described above. In this unbalanced region, the outer circulation stimulates planetary Rossby waves, distorting the flow and transferring kinetic energy out of the vortex.

%


\acknowledgments

The authors thank Malte Jansen for suggesting testing the vortex Rhines.timescale. Funding support was provided by NSF grants 1826161 and 1945113.




%
%
%


 \bibliographystyle{ametsoc2014}
 \bibliography{references}

\end{document}